# Simultaneous Evaluation of Heat Capacity and In-plane Thermal Conductivity of Nanocrystalline Diamond Thin Films

Luke Yates,[1] Zhe Cheng,[1] Tingyu Bai,[2] Karl Hobart,[3] Marko Tadjer,[3] Tatyana I. Feygelson,[3] Bradford B. Pate,[3] Mark Goorsky,[2] Samuel Graham,[1, a)]

[1]George W. Woodruff School of Mechanical Engineering, Georgia Institute of Technology, Atlanta, 30332, USA

[2]Materials Science and Engineering, University of California, Los Angeles, Los Angeles, CA, 91355, USA

[3]U.S. Naval Research Laboratory, 4555 Overlook Ave SW, Washington, DC, 20375, USA

[a)] Corresponding author: sgraham@gatech.edu

**Abstract**

As wide bandgap electronic devices have continued to advance in both size reduction and power handling capabilities, heat dissipation has become a significant concern. To mitigate this, chemical vapor deposited (CVD) diamond has been demonstrated as an effective solution for thermal management of these devices by directly growing onto the transistor substrate. A key aspect of power and radio frequency (RF) electronic devices involves transient switching behavior, which highlights the importance of understanding the temperature dependence of a material's heat capacity and thermal conductivity when modeling and predicting device electro-thermal response. Due to the complicated microstructure near the interface between CVD diamond and electronics, it is difficult to measure both properties simultaneously. In this work, we use time-domain thermoreflectance (TDTR) to simultaneously measure the in-plane thermal conductivity and heat capacity of a 1-µm-thick CVD diamond film, and also use the pump as an effective heater to perform temperature dependent measurements. The results show that the in-plane thermal conductivity varied slightly with an average of 103 W/m-K over a temperature range of 302-327 K, while the specific heat capacity has a strong temperature dependence over the same range and matches with heat capacity data of natural diamond in literature.

**Introduction**

The exceptional properties of diamond have long been investigated in an effort to take advantage of this unique material. Single crystal diamond has the highest thermal conductivity of any known three-dimensional solid, up to 2500 W/m-K at room temperature.[1] In addition, diamond is a dielectric material making it well-suited to act as an excellent heat spreader in high power electronic devices.[2-4] Because of the difficulty of integrating natural diamond into electronic devices, synthetic diamond grown through CVD and other techniques has been used since the 1980s.[5] The nature of the diamond growth allows for an environment where both nano and micro-crystalline material can be grown, as well as bulk material with very large grains and thermal conductivities approach their natural counterpart.[6-8] Therefore, significant effort has gone into fully characterizing synthetic diamond of all varieties.[9-10]

In terms of power electronic and radio frequency (RF) devices, CVD diamond has been proposed to integrate with GaN-based high-electron-mobility transistors (HEMTs) as both a heat spreading layer and a device substrate.[2, 11-14] The goal is to significantly improve the heat dissipation of the devices and to decrease the operational temperature. It has been shown that limiting device operational temperature can significantly increase the reliability and lifetime of the device.[15] Both power electronic and RF electronics applications require that the transistors operate under a pulsed condition ranging from a few KHz in power electronics to several GHz in RF amplifiers.[16] This highlights the importance of understanding the temperature dependence of both the diamond heat capacity and the thermal conductivity when designing and modeling devices.

Moelle *et al.* used a differential scanning calorimeter (DSC) to measure the specific heat of a 300-nm-thick nanocrystalline diamond sample and a 300-µm-thick microcrystalline diamond sample as a function of temperature. They found excellent agreement between the single crystal reference sample and the microcrystalline sample, while the nanocrystalline sample did show some deviation in the specific heat when comparing to the single crystalline and the microcrystalline samples.[9] The recent achievements in CVD diamond growth revealed the importance of seeding density and especially seed distribution uniformity to the deposition of higher quality NCD films. Ultrasonic treatment of a substrate in nanodiamond suspension became a standard seeding technique that can provide for seeding density over $10^{12}$ nuclei/cm$^2$, uniform distribution of diamond seed, as well as to avoid damage to substrate structure. A low methane to hydrogen ratio during diamond CVD deposition is one of the conditions that prevents secondary renucleation and ensures that NCD films exhibit a pronounced columnar grain texture resulting only from a competitive growth of original randomly oriented diamond seeds. Since the growth of CVD diamond onto electronic material, such as GaN, introduces a nanocrystalline structure that is responsible for an increase in thermal resistance as compared to its bulk counterpart, it is important to characterize the thermal conductivity and heat capacity of these layers near the growth interface.

In this work, we utilized time-domain thermoreflectance (TDTR) to simultaneously measure the in-plane thermal conductivity and the heat capacity of a 1-µm-thick suspended CVD nanocrystalline diamond membrane. The measurements were performed at five power conditions to introduce heating in the diamond membranes. For each power condition, multi-frequency TDTR was used to obtain different TDTR sensitivities of the heat capacity and the in-plane thermal conductivity. The temperature rise due to laser heating in the membrane was then calculated to measure the temperature dependence of both the heat capacity and the in-plane thermal conductivity simultaneously.

## Sample and Methods

A 1-µm-thick suspended diamond membrane grown via CVD was used to allow for simultaneous measurements of both in-plane thermal conductivity and heat capacity. First, a tensile diamond film was grown on a silicon substrate using 0.3% methane to hydrogen ratio at a temperature of 750°C with a growth pressure of 7 Torr and the microwave power was varied from 800 W to 1400 W during growth. Additional growth details can be found in the following reference.[17] Then through standard photolithography, a square region (3000 µm x 3000 µm) of the silicon substrate was isolated and etched away, leaving a 1-µm-thick suspended membrane of CVD diamond. A schematic of the sample structure is shown in Figure 1.

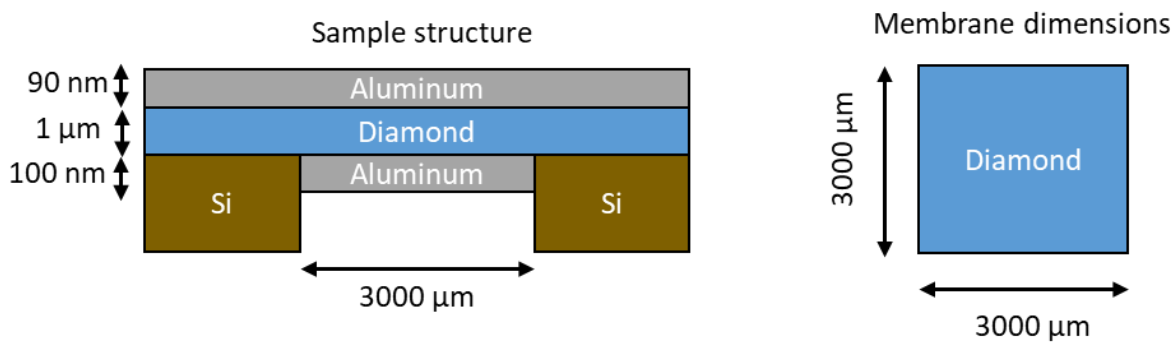

**Figure 1**. Schematic diagram of the sample structure. A 300 µm x 300 µm 1-µm-thick diamond membrane was fabricated by etching the Si on the backside. Subsequently, ~90 nm of Al was depositied on the top and ~100 nm of Al was deposited on the bottom of the diamond to act as TDTR transducer.

According to the transmission electron microscopy (TEM) measurements, the sample was found to have an average grain size of 135 nm near the growth surface with a maximum and minimum of 267 nm and 63 nm, respectively. Near the nucleation interface, the average grain size was 97 nm with a maximum and minimum of 198 nm and 27 nm, respectively. The root-mean-square (RMS) surface roughness was measured by atomic force microscopy (AFM) to be 19 ± 2 nm.[18] The cross section of the suspended diamond layer and grain size information are shown in Figure 2. The suspended diamond membrane forces heat to flow laterally, thereby enhancing the TDTR sensitivity to measure the in-plane thermal conductivity.

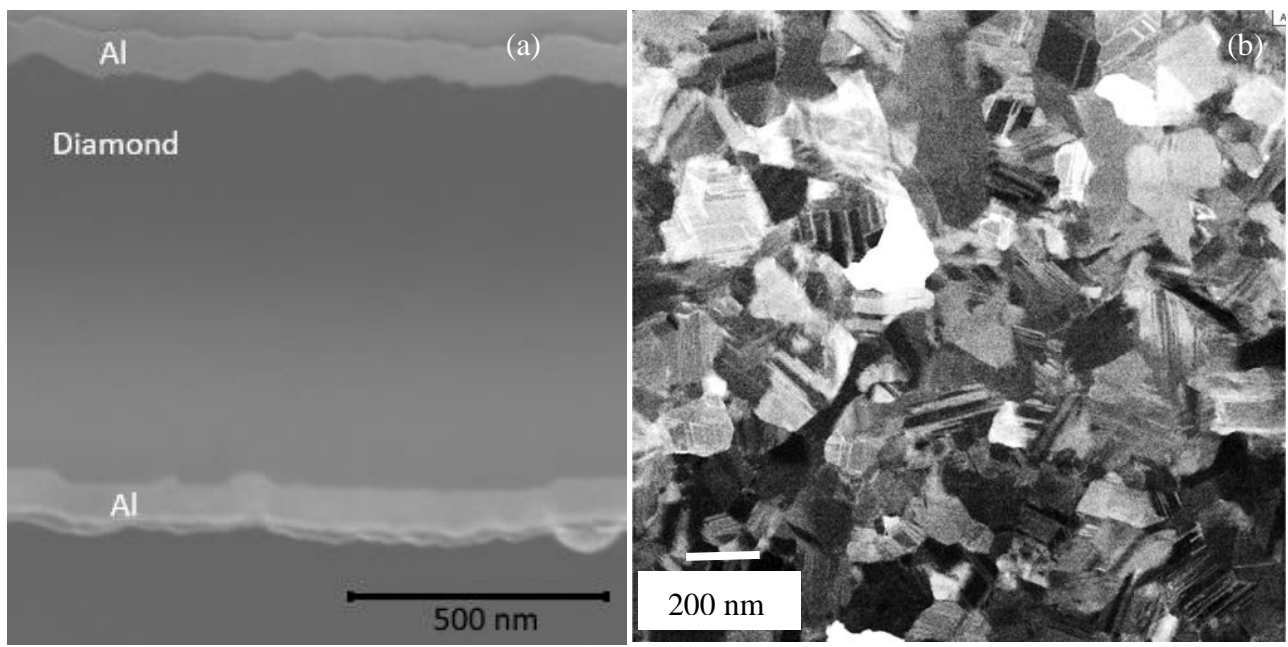

**Figure 2**. (a) SEM image of a 1-µm-thick suspended diamond membrane with Al deposited on both top and bottom sides. (b) The grain size distribution of the sample near the top growth surface.

TDTR was used to measure the thermal properties of the CVD diamond membrane as well as to heat the diamond membrane by the pump beam which enables temperature-dependent measurements of specific heat. TDTR is an optical pump-probe method to measure thermal properties of both bulk and nanostructures with an ultrafast laser.[19-21] Polarizing optics are then used to split the laser into two beams (pump and probe). The pump beam is further modulated using an electro-optic modulator (EOM). The modulated pump beam heats the sample surface periodically while the probe beam monitors the temperature of the sample surface via the change of reflectivity after passing through a mechanical delay stage. By moving the delay stage, the probe pulse is delayed with picosecond time resolution relative to the pump. The full travel of the delay stage allows for the probe to infer the thermal decay of the pump pulses on the sample surface. The sample is coated with a thin metal transducer which creates a known heat flux on the sample surface and has a large coefficient of thermoreflectance at the probe wavelength which improves the overall signal-to-noise ratio. The probe beam is reflected back from the sample surface and directed to a photodiode that is connected to a lock-in amplifier in sync with the EOM. The signal picked up by the lock-in amplifier is then used to infer unknown thermal parameters by fitting to an analytical heat transfer solution.[19, 21] More details about the system used in this work can be found at the following reference.[22]

For TDTR measurements, thin films are usually supported by a substrate that allows for effective heat dissipation, thereby keeping the temperature rise within a few degrees during the measurements.[23] However, in our work, the 1-µm-thick thin film is suspended in air which significantly impedes heat conduction in the through-plane direction. The surface temperature of a multi-layered structure under periodic laser heating has been reported in the literature.[19-21] We used these models to estimate the steady-state temperature rise in our measurements for a given pump diameter and incident power. For a pump beam with a 20-µm diameter and a power of 10 mW, Figure 3 shows the calculated steady-state temperature rise for both the membrane and the supported film. It is clear to see that the substrate provides an extra path for heat dissipation as compared to the suspended membrane. The membrane geometry allows for significant heating and subsequent temperature-dependent thermal measurements.

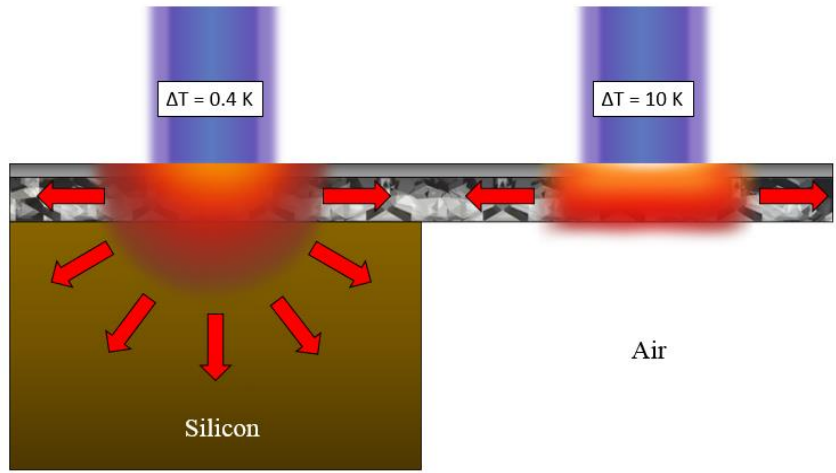

**Figure 3**. Schematic diagram of how the laser energy from the TDTR pump diffuses through the material. When the diamond film is supported on Si, a steady-state temperature rise of < 0.5 K is obtained. However, since the heat is forced to transfer laterally on the membrane, a large steady-state temperature rise of 10 K is obtained for the same 10 mW power condition.

In addition to the capability of heating the sample by the pump beam, the modulation frequencies of the pump beam were varied to separate the in-plane thermal conductivity and the heat capacity of the diamond membrane. Multi-frequency TDTR has been used to simultaneously measure thermal conductivity and heat capacity of supported thin films before.[21] Here, by increasing the pump power rather than using an external heating stage, we show an additional ability to evaluate the temperature dependence of thermal properties. In these measurements, we used a 10x objective with pump and probe spot diameters of 20 ± 0.5 µm and 12.4 ± 0.5 µm, respectively. The pump and probe powers were measured before the objective. The transmission through the objective for pump and probe are 0.7 and 0.6, respectively. The reflectivity of the Al transducer at the pump and probe wavelengths (400 nm and 800 nm) are measured to be 0.90 and 0.85. The TDTR measurements were carried out at five different pump powers while the probe power was kept constant (2.8 mW). The pump power was set to 10.5, 20.3, 30.2, 40.3, and 50.3 mW before the objective. The absorbed pump power by the Al transducer was 0.735, 1.421, 2.114, and 3.521 mW while the absorbed probe power was 0.09 mW. At each power condition, three frequencies were used for the pump modulation: 1.2, 3.6, and 6.3 MHz. The range of frequencies were selected to provide enough sensitivity to the parameters of interest, which will be discussed later. All the measurements were performed on the same sample spot to avoid the effect of variation of sample structure. [24]

## Results and Discussion

The derivation of spatial and temporal temperature distributions from a modulated heat source on multi-layered structures has been reported in literature.[25] In this work, we define the temperature rise of the sample as a weighted average steady-state temperature rise across the pump beam. We only consider the steady-state temperature rise so the modulation frequency does not affect our calculation. As shown in Figure 4 (left axis), the 20-µm-diameter pump beam (10.5 mW before objective) heats up the > 100 µm-diameter area of the diamond membrane, showing the heat spreading effect. The max temperature rise induced by the pump beam (~7 K) is at the center of the beam while the weighted average temperature rise is ~6.5 K. The right axis of Figure 4 shows the normalized intensity distribution of the probe beam which is narrowly focused on the center, which induces an additional temperature rise of ~0.8 K. Therefore, the overall steady-state temperature rise is ~7.3 K. Room temperature was measured to be 296 K to calculate the absolute temperature.

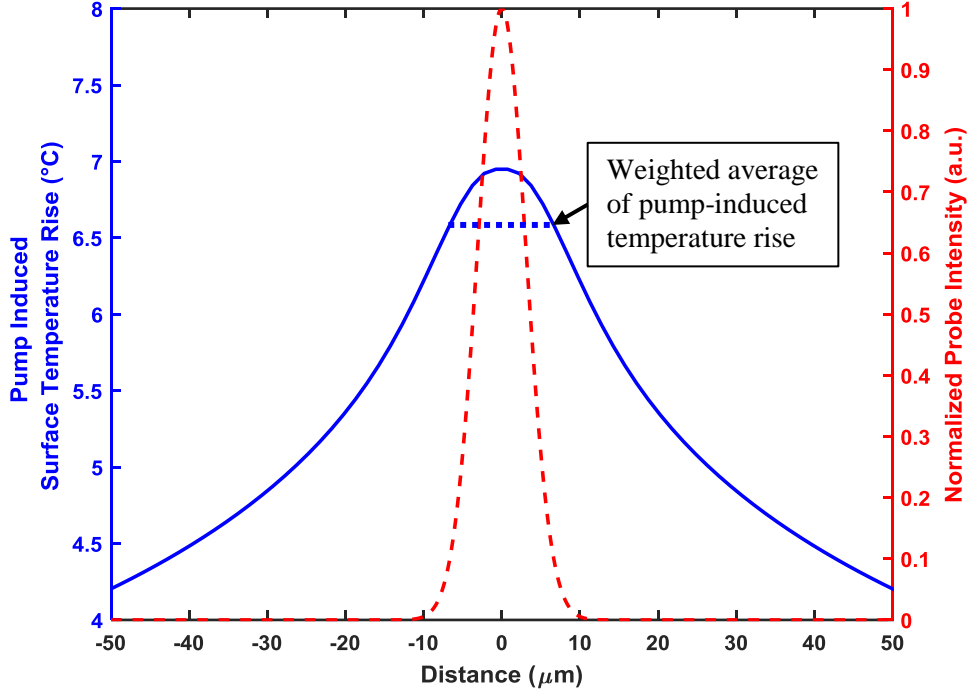

**Figure 4**. The temperature rise distribution induced by the pump beam (left axis) and the probe beam intensity distribution (right-axis).

The sensitivity of each parameter is defined as a fractional change in the measured signal to an individual parameter $p$:

$$S_p = \frac{d(\ln R)}{(\ln p)} = \frac{dR/R}{dp/p} \quad (1)$$

where $R$ is the measured ratio of the in-phase signal to the out-of-phase signal ($-V_{in}/V_{out}$). It is notable that the positive and negative values only indicate how that parameter changes the direction of the curve, and only the absolute value of sensitivity matters.[26] For the sample studied in this work, there were three unknown parameters of interest: the diamond heat capacity, the diamond in-plane thermal conductivity, and the thermal boundary conductance (TBC) between the Al transducer and the diamond membrane. For the multi-frequency TDTR technique, it is desirable to have large differences in the sensitivity of individual parameters at a certain frequency.[21] As shown in Figure 5, the sensitivity of the diamond specific heat decreases with modulation frequency while that of the in-plane thermal conductivity increases. Therefore, by including both 1.2 MHz and 6.3 MHz in the data fitting, we were able to effectively separate both parameters. The last unknown parameter, the Al/diamond TBC was significant at all frequencies. Additional parameters of interest include the Al transducer thickness which is measured using a well-documented picosecond acoustics method [27-28], and the through-plane diamond thermal conductivity. The exact parameters used in the sensitivity analysis are listed in Table 1. The Al thermal conductivity was determined by measuring electrical conductivity and applying the Wiedemann-Franz law.

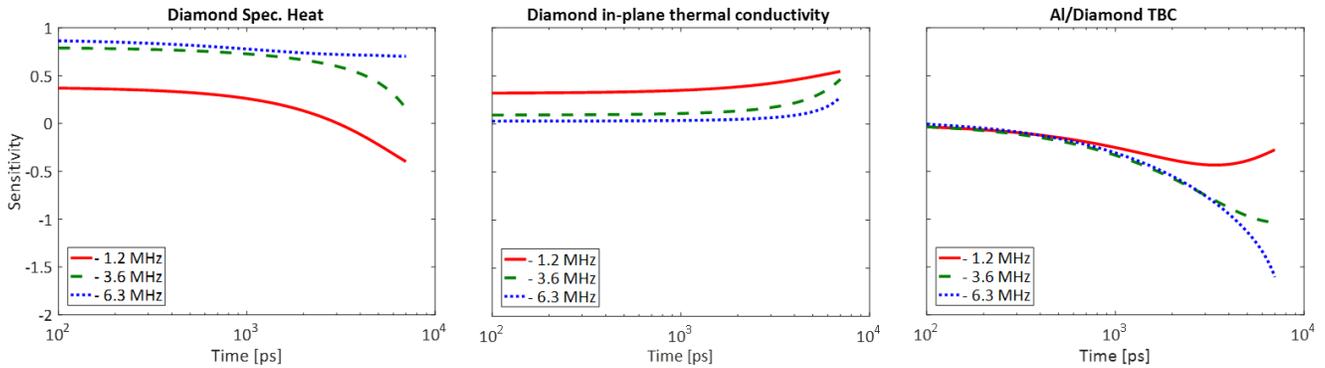

**Figure 5**. Sensitivity plots for the unknown parameters at three frequencies. Since the sensitivity of the heat capacity and the in-plane thermal conductivity are distinctly different at low and high modulation frequencies, we were able to fit for both parameters simultaneously.

The CVD diamond has been well-documented to have an inherent gradient microstructure through the thickness due to the columnar grain growth.[29-31] The through-plane thermal conductivity value used here was measured on a spot where the diamond is supported by the silicon substrate. The TDTR measurement was taken at 11.6 MHz with a 5x objective to induce near 1D heat transfer through the diamond. We obtained an effective value of 175 +65/-42 W/m-K for the through-plane diamond thermal conductivity [17], and is consistent with the reported values in literature.[24, 32-33] The large error results from the small sensitivity of the through-plane thermal conductivity. This value along with its uncertainty was used in the data fitting of the suspended membrane as a known parameter. Because of the geometry of the suspended membrane, in-plane heat conduction dominates during the measurements so the sensitivity of the through-plane thermal conductivity is very small compared with that of the in-plane thermal conductivity and the heat capacity. Therefore, the error in the through-plane thermal conductivity has a negligible effect on the measurements of the in-plane thermal conductivity and the heat capacity.

**Table 1**. Material properties used in the TDTR sensitivity analysis and data fitting. $C_{p\text{-diamond}}$, $k_{r\text{-diamond}}$, and $G_{\text{diamond}}$ are unknown parameters.

|  | $k_z$ [W/m-K] | $k_r$ [W/m-k] | $\rho$ [kg/m$^3$] | $c_p$ [J/kg-K] | d [nm] | G [MW/m$^2$-K] |
|---|---|---|---|---|---|---|
| Diamond | 175 +65/-42 | fit | 3500 | fit | 1000 ± 50 | fit |
| Al | 175 ± 35 | 175 | 2700 | 900 ± 45 | 87 ± 3* | |

The fitting of the experimental data with the analytical solution is shown in Figure 6 and excellent agreement is achieved for all the power conditions. The error bars were calculated by a Monte Carlo method in which we assigned an uncertainty to each parameter in the model and those parameters were randomly varied within the uncertainty bounds before fitting for the three unknown parameters. This is repeated 500 times to develop a distribution. The 50$^{th}$ percentile value is taken as the measured value with the 90$^{th}$ and 10$^{th}$ percentile being used as the upper and lower error bounds.[22] The measured diamond heat capacity are shown in Figure 7. We also included literature values of natural diamond as comparison.[34-35]

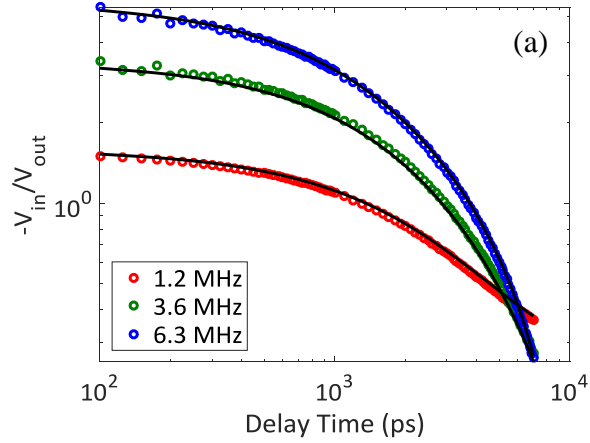

**Figure 6**. Data fitting of the experimental data (circles) to the analytical model (lines) for three frequencies simultaneously.

Both theoretical and experimental methods have demonstrated deviations of diamond heat capacity as a function of grain size.[9, 36] These variations were attributed to changes in the phonon dispersion as a result of characteristic length scales reduction and phonon softening at grain boundaries. Theoretically, heat capacity should increase as the characteristic length scale for thermal energy transport in the material is reduced.[36] Additionally, grain boundaries and growth concerns such as non-diamond carbon content, cracking, and chemical impurities may change both the density and heat capacity.[29] In this work, the 1-µm-thick suspended diamond membrane was measured to have a similar heat capacity to natural diamond, which is a result of all the factors mentioned above.

Accurate measurements of both specific heat and thermal conductivity are of importance when utilizing diamond to extract heat from power electronic devices and RF devices. Diamond has one of the highest Debye temperatures of any known material which was reported to be 1800 K - 2200 K.[34, 37] Its heat capacity has a strong temperature dependence below the Debye temperature and will increase significantly at room temperature and above, as shown in Figure 7. The vertical error bars in Figure 7 are calculated by the Monte Carlo Method. The horizontal error bars are the highest and lowest calculated temperatures when varying the properties in the model within the associated parameter uncertainties in Table 1.

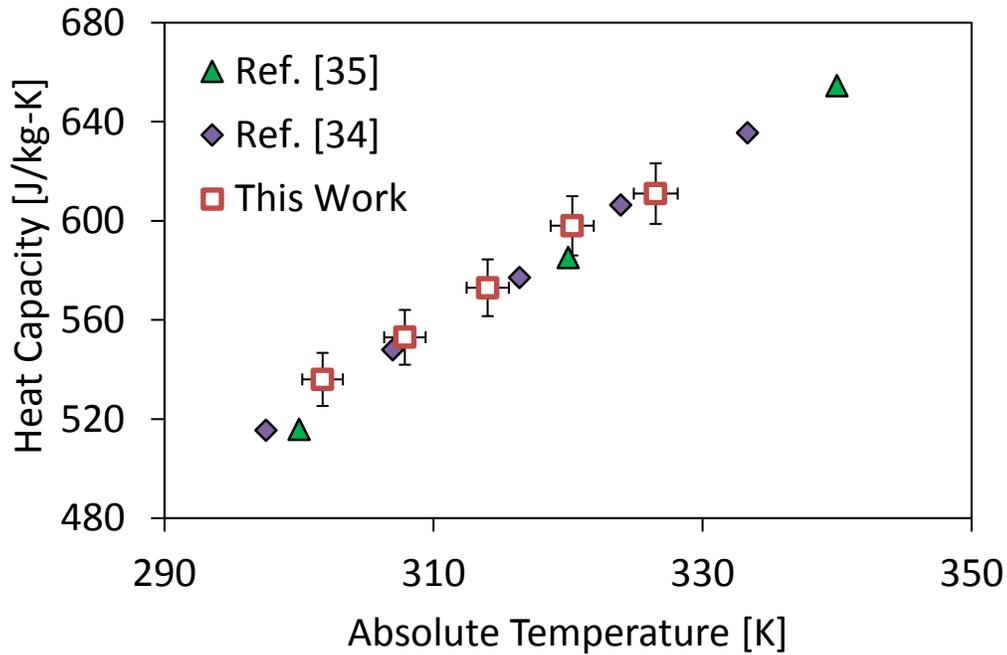

Figure 7: The heat capacity of the nanocrystalline diamond membrane measured by TDTR in this work. As comparison, the measured heat capacity data of natural diamond from literature are also included.[34-35]

For the measured in-plane thermal conductivity, our results are consistent with previous reports on the measured in-plane thermal conductivity of CVD diamond films which showed weak temperature dependence within 298-498 K.[29, 33, 38] The measured in-plane thermal conductivity in this work did not show obvious temperature dependence. Table 2 lists the values of the measured values for all the power conditions. The in-plane thermal conductivity showed a slight variation from 92 W/m-K to 115 W/m-K with an average of 103 ± 12 W/m-K.[17, 24] Additionally, there is a slight change in the Al/diamond TBC from each power condition with a high and low value of 214 ± 7 MW/m$^2$K and 222 ± 6 MW/m$^2$K, respectively. It has been shown by Monachon *et al.* and others that the Al/diamond TBC can vary significantly depending on the surface chemistry and growth conditions. They have reported this value to be as low as 32 MW/m²K and up to 230 MW/m²K by simply changing the surface treatment prior to the metal depositions.[39-41]

**Table 2**. Measured values for the unknown parameters as related to the pump heating power and the temperature rise in the diamond film.

| Power Before Objective | Total Incident Power at Sample Surface | $\Delta T$ | $k_r$ | $C_p$ | TBC $_{Al-dia}$ |
|---|---|---|---|---|---|
| mW | mW | K | W/m-K | J/kg-K | MW/m²K |
| 10.5 | 0.83 | 7.3 | 99 | 536 | 214 |
| 20.3 | 1.51 | 13.4 | 92 | 553 | 214 |
| 30.2 | 2.20 | 19.6 | 106 | 573 | 219 |
| 40.3 | 2.91 | 25.9 | 103 | 598 | 218 |
| 50.3 | 3.61 | 32.1 | 115 | 611 | 222 |

## Conclusions

In this work, we used TDTR to simultaneously measure the in-plane thermal conductivity and the heat capacity of a CVD nanocrystalline diamond thin film. The suspended membrane structure allowed us to induce significant laser heating from the TDTR pump beam. By systematically increasing the pump power in the TDTR experiment, we demonstrated the ability to effectively heat the sample to study temperature dependence of the thermal properties. The results show that the in-plane thermal conductivity varied slightly with an average of 103 W/m-K over a temperature range of 302-327 K, while the heat capacity of the diamond showed a strong temperature dependence. The heat capacity was measured to be 536 J/kg-K at 302 K, 553 J/kg-K at 308 K, 573 J/kg-K at 314 K, 598 J/kg-K at 320 K, and 611 J/kg-K at 327 K. A good understanding of the temperature dependence of both the thermal conductivity and the heat capacity is crucial to provide accurate device modeling when integrating CVD diamond into both power electronic and RF devices.

## Acknowledgments

This material is based upon work supported by the National Science Foundation Graduate Research Fellowship Program under Grant No. DGE-1650044. We also acknowledge support of DARPA: Thermal Transport in Diamond Thin Films for Electronic Thermal Management initiative under contract no. FA8650-15C-7517. Any opinions, findings, and conclusions or recommendations expressed in this material are those of the authors and do not necessarily reflect the views of the National Science Foundation or DARPA.

**References**

(1) Graebner, J. E. Thermal conductivity of diamond. In *Diamond: Electronic properties and applications*; Springer: 1995; pp 285-318.
(2) Altman, D.; Tyhach, M.; McClymonds, J.; Kim, S.; Graham, S.; Cho, J.; Goodson, K.; Francis, D.; Faili, F.; Ejeckam, F. In *Analysis and characterization of thermal transport in GaN HEMTs on Diamond substrates*, Thermal and Thermomechanical Phenomena in Electronic Systems (ITherm), 2014 IEEE Intersociety Conference on, IEEE: 2014; pp 1199-1205.
(3) Jagannadham, K. Multilayer diamond heat spreaders for electronic power devices. *Solid-State Electronics* **1998,** *42* (12), 2199-2208.
(4) Han, Y.; Lau, B. L.; Tang, G.; Zhang, X. Thermal management of hotspots using diamond heat spreader on Si microcooler for GaN devices. *IEEE Transactions on Components, Packaging and Manufacturing Technology* **2015,** *5* (12), 1740-1746.
(5) Kobashi, K.; Nishimura, K.; Kawate, Y.; Horiuchi, T. Synthesis of diamonds by use of microwave plasma chemical-vapor deposition: Morphology and growth of diamond films. *Physical review B* **1988,** *38* (6), 4067.
(6) Yates, L.; Cheaito, R.; Sood, A.; Cheng, Z.; Bougher, T.; Asheghi, M.; Goodson, K.; Goorsky, M.; Faili, F.; Twitchen, D. In *Investigation of the Heterogeneous Thermal Conductivity in Bulk CVD Diamond for Use in Electronics Thermal Management*, ASME 2017 International Technical Conference and Exhibition on Packaging and Integration of Electronic and Photonic Microsystems collocated with the ASME 2017 Conference on Information Storage and Processing Systems, American Society of Mechanical Engineers: 2017; pp V001T04A014-V001T04A014.
(7) Twitchen, D.; Pickles, C.; Coe, S.; Sussmann, R.; Hall, C. Thermal conductivity measurements on CVD diamond. *Diamond and related materials* **2001,** *10* (3-7), 731-735.
(8) Graebner, J.; Altmann, H.; Balzaretti, N.; Campbell, R.; Chae, H.-B.; Degiovanni, A.; Enck, R.; Feldman, A.; Fournier, D.; Fricke, J. Report on a second round robin measurement of the thermal conductivity of CVD diamond. *Diamond and Related materials* **1998,** *7* (11-12), 1589-1604.
(9) Moelle, C.; Werner, M.; SzuÈcs, F.; Wittorf, D.; Sellschopp, M.; Von Borany, J.; Fecht, H.-J.; Johnston, C. Specific heat of single-, poly-and nanocrystalline diamond. *Diamond and related materials* **1998,** *7* (2-5), 499-503.
(10) Lani, S.; Ataman, C.; Noell, W.; Briand, D.; de Rooij, N. In *Thermal characterization of polycrystalline CVD diamond thin films*, Proceedings of Eurosensors 2009, Elsevier Science, Reg Sales Off, Customer Support Dept, 655 Ave Of The Americas, New York, Ny 10010 Usa: 2009; p 108.

(11) Diduck, Q.; Felbinger, J.; Eastman, L.; Francis, D.; Wasserbauer, J.; Faili, F.; Babic, D. I.; Ejeckam, F. Frequency performance enhancement of AlGaN/GaN HEMTs on diamond. *Electronics letters* **2009,** *45* (14), 758-759.
(12) Jessen, G.; Gillespie, J.; Via, G.; Crespo, A.; Langley, D.; Wasserbauer, J.; Faili, F.; Francis, D.; Babic, D.; Ejeckam, F. In *AlGaN/GaN HEMT on diamond technology demonstration*, Compound Semiconductor Integrated Circuit Symposium Proceedings, 2006; pp 271-274.
(13) Zhang, R.; Zhao, W.; Yin, W.; Zhao, Z.; Zhou, H. Impacts of diamond heat spreader on the thermo-mechanical characteristics of high-power AlGaN/GaN HEMTs. *Diamond and Related Materials* **2015,** *52*, 25-31.
(14) Anaya, J.; Sun, H.; Pomeroy, J.; Kuball, M. In *Thermal management of GaN-on-diamond high electron mobility transistors: Effect of the nanostructure in the diamond near nucleation region*, Thermal and Thermomechanical Phenomena in Electronic Systems (ITherm), 2016 15th IEEE Intersociety Conference on, IEEE: 2016; pp 1558-1565.
(15) Smith, K.; Barr, R., Reliability lifecycle of GaN power devices. 2016.
(16) Amano, H.; Baines, Y.; Beam, E.; Borga, M.; Bouchet, T.; Chalker, P. R.; Charles, M.; Chen, K. J.; Chowdhury, N.; Chu, R. The 2018 GaN power electronics roadmap. *Journal of Physics D: Applied Physics* **2018,** *51* (16), 163001.
(17) Anaya, J.; Bai, T.; Wang, Y.; Li, C.; Goorsky, M.; Bougher, T.; Yates, L.; Cheng, Z.; Graham, S.; Hobart, K. Simultaneous determination of the lattice thermal conductivity and grain/grain thermal resistance in polycrystalline diamond. *Acta Materialia* **2017,** *139*, 215-225.
(18) Nečas, D.; Klapetek, P. Gwyddion: an open-source software for SPM data analysis. *Open Physics* **2012,** *10* (1), 181-188.
(19) Cahill, D. G. Analysis of heat flow in layered structures for time-domain thermoreflectance. *Review of scientific instruments* **2004,** *75* (12), 5119-5122.
(20) Schmidt, A. J.; Chen, X.; Chen, G. Pulse accumulation, radial heat conduction, and anisotropic thermal conductivity in pump-probe transient thermoreflectance. *Review of Scientific Instruments* **2008,** *79* (11), 114902.
(21) Liu, J.; Zhu, J.; Tian, M.; Gu, X.; Schmidt, A.; Yang, R. Simultaneous measurement of thermal conductivity and heat capacity of bulk and thin film materials using frequency-dependent transient thermoreflectance method. *Review of Scientific Instruments* **2013,** *84* (3), 034902.
(22) Bougher, T. L.; Yates, L.; Lo, C.-F.; Johnson, W.; Graham, S.; Cola, B. A. Thermal boundary resistance in GaN films measured by time domain thermoreflectance with robust Monte Carlo uncertainty estimation. *Nanoscale and Microscale Thermophysical Engineering* **2016,** *20* (1), 22-32.
(23) Schmidt, A. J. Optical characterization of thermal transport from the nanoscale to the macroscale. Massachusetts Institute of Technology, 2008.
(24) Cheaito, R.; Sood, A.; Yates, L.; Bougher, T. L.; Cheng, Z.; Asheghi, M.; Graham, S.; Goodson, K. In *Thermal conductivity measurements on suspended diamond membranes using picosecond and femtosecond time-domain thermoreflectance*, Thermal and Thermomechanical Phenomena in Electronic Systems (ITherm), 2017 16th IEEE Intersociety Conference on, IEEE: 2017; pp 706-710.
(25) Braun, J. L.; Szwejkowski, C. J.; Giri, A.; Hopkins, P. E. On the steady-state temperature rise during laser heating of multilayer thin films in optical pump–probe techniques. *Journal of Heat Transfer* **2018,** *140* (5).
(26) Hamby, D. A review of techniques for parameter sensitivity analysis of environmental models. *Environmental monitoring and assessment* **1994,** *32* (2), 135-154.
(27) Hohensee, G. T.; Hsieh, W.-P.; Losego, M. D.; Cahill, D. G. Interpreting picosecond acoustics in the case of low interface stiffness. *Review of Scientific Instruments* **2012,** *83* (11), 114902.
(28) Antonelli, G. A.; Perrin, B.; Daly, B. C.; Cahill, D. G. Characterization of mechanical and thermal properties using ultrafast optical metrology. *MRS bulletin* **2006,** *31* (8), 607-613.

(29) Graebner, J.; Jin, S.; Kammlott, G.; Bacon, B.; Seibles, L.; Banholzer, W. Anisotropic thermal conductivity in chemical vapor deposition diamond. *Journal of applied physics* **1992,** *71* (11), 5353-5356.
(30) Cheng, Z.; Bougher, T.; Bai, T.; Wang, S. Y.; Li, C.; Yates, L.; Foley, B. M.; Goorsky, M.; Cola, B. A.; Faili, F. Probing Growth-Induced Anisotropic Thermal Transport in High-Quality CVD Diamond Membranes by Multifrequency and Multiple-Spot-Size Time-Domain Thermoreflectance. *ACS applied materials & interfaces* **2018,** *10* (5), 4808-4815.
(31) Cheng, Z.; Foley, B. M.; Bougher, T.; Yates, L.; Cola, B. A.; Graham, S. Thermal rectification in thin films driven by gradient grain microstructure. *Journal of Applied Physics* **2018,** *123* (9), 095114.
(32) Anaya, J.; Rossi, S.; Alomari, M.; Kohn, E.; Tóth, L.; Pécz, B.; Kuball, M. Thermal conductivity of ultrathin nano-crystalline diamond films determined by Raman thermography assisted by silicon nanowires. *Applied Physics Letters* **2015,** *106* (22), 223101.
(33) Anaya, J.; Rossi, S.; Alomari, M.; Kohn, E.; Tóth, L.; Pécz, B.; Hobart, K. D.; Anderson, T. J.; Feygelson, T. I.; Pate, B. B. Control of the in-plane thermal conductivity of ultra-thin nanocrystalline diamond films through the grain and grain boundary properties. *Acta Materialia* **2016,** *103*, 141-152.
(34) Victor, A. C. Heat capacity of diamond at high temperatures. *The Journal of Chemical Physics* **1962,** *36* (7), 1903-1911.
(35) Raman, C. In *The heat capacity of diamond between 0 and 1000° K*, Proceedings of the Indian Academy of Sciences-Section A, Springer: 1957; pp 323-332.
(36) Adiga, S. P.; Adiga, V. P.; Carpick, R. W.; Brenner, D. W. Vibrational properties and specific heat of ultrananocrystalline diamond: Molecular dynamics simulations. *The Journal of Physical Chemistry C* **2011,** *115* (44), 21691-21699.
(37) Spear, K. E.; Dismukes, J. P. *Synthetic diamond: emerging CVD science and technology*, John Wiley & Sons: 1994; Vol. 25.
(38) Zhou, Y.; Ramaneti, R.; Anaya, J.; Korneychuk, S.; Derluyn, J.; Sun, H.; Pomeroy, J.; Verbeeck, J.; Haenen, K.; Kuball, M. Thermal characterization of polycrystalline diamond thin film heat spreaders grown on GaN HEMTs. *Applied Physics Letters* **2017,** *111* (4), 041901.
(39) Monachon, C.; Weber, L. Effect of diamond surface orientation on the thermal boundary conductance between diamond and aluminum. *Diamond and Related Materials* **2013,** *39*, 8-13.
(40) Collins, K. C.; Chen, S.; Chen, G. Effects of surface chemistry on thermal conductance at aluminum–diamond interfaces. *Applied Physics Letters* **2010,** *97* (8), 083102.
(41) Hohensee, G. T.; Wilson, R.; Cahill, D. G. Thermal conductance of metal–diamond interfaces at high pressure. *Nature communications* **2015,** *6*, 6578.